\documentstyle[12pt]{article}
\def\be{\begin{equation}}
\def\ee{\end{equation}}
\newcommand{\tr}{\mbox{tr}}
\newcommand{\sutwo}{\mbox{SU}(2)}
\newcommand{\suthree}{\mbox{SU}(3)}
\newcommand{\sufour}{\mbox{SU}(4)}
\newcommand{\uone}{\mbox{U}(1)}
\newcommand{\suN}{\mbox{SU}(N)}
\newcommand{\suNM}{\mbox{SU}(N+M)}
\newcommand{\bi}{\bibitem}
\begin{document}
\baselineskip=15.5pt

\begin{titlepage}

\begin{flushright}
hep-th/0101020\\
PUPT-1974\\
\end{flushright}
\vfil

\begin{center}
{\huge Gravity Duals of Fractional Branes in Various Dimensions}\\
\vspace{3mm}
\end{center}

\vfil
\begin{center}
{\large Christopher P. Herzog and Igor R. Klebanov}\\
\vspace{1mm}
Joseph Henry Laboratories, Princeton University,\\
Princeton, New Jersey 08544, USA\\
\vspace{3mm}
\end{center}

\vfil

\begin{center}
{\large Abstract}
\end{center}

\noindent
We derive type II supergravity solutions
corresponding to space-filling regular and fractional D$p$ branes 
on $(9-p)$-dimensional conical transverse spaces.
Fractional D$p$-branes are wrapped D$(p+2)$-branes; therefore,
our solutions exist only if the base of the cone has a non-vanishing
Betti number $b_2$.  
We also consider 11-dimensional SUGRA solutions corresponding to
regular and fractional M2 branes on 8-dimensional cones 
whose base has a non-vanishing $b_3$. (In this case a fractional
M2-brane is an M5-brane wrapped over a 3-cycle.)
We discuss the gauge theory intepretation of these solutions, as well
as of the solutions constructed by Cveti\v{c} et al. in hep-th/0011023 and
hep-th/0012011.
\vfil
\begin{flushleft}
December 2000
\end{flushleft}
\vfil
\end{titlepage}
\newpage
\renewcommand{\baselinestretch}{1.1}  %looks better

%%%%%%%%%%%%%%%%%%%%%%%%%%%%%%%%%%%%%%%%%%%%%
%% include the next line for double spacing %%
%%%%%%%%%%%%%%%%%%%%%%%%%%%%%%%%%%%%%%%%%%%%%%
%\renewcommand{\baselinestretch}{2}

\section{Introduction}

The basic AdS/CFT correspondence 
\cite{jthroat,US,EW} is motivated by comparing a stack of 
elementary branes
with the metric it produces (for reviews, see for example
\cite{magoo,me}). In order to break some of the
supersymmetry, we may place the stack
at a conical singularity \cite{ks,Kehag,KW,MP}.
Consider, for instance, a stack of D3-branes placed at the apex of a 
Ricci-flat 6-d cone $Y_6$ whose base is a 5-d Einstein manifold
$X_5$. Comparing the metric with the D-brane description leads
one to conjecture that type IIB string theory on $AdS_5\times X_5$
is dual to the low-energy limit of the world volume theory on
the D3-branes at the singularity.

For certain cones $Y_6$, in addition to the regular D3-branes
which may be moved away from the apex, there are
also fractional D3-branes which can exist only at the 
singularity \cite{gipol,doug,GK,KN}. 
These fractional branes may be thought of as D5-branes wrapped over
2-cycles of $X_5$, the base of the cone. If we are interested in
stacking arbitrary numbers of such branes, we have to require that
the second homology group $H_2(X_5)$ is ${\bf Z}$ or bigger.

An example of a smooth $X_5$ with Betti number $b_2=1$ is the Einstein space
$T^{1,1}$ which is topologically $S^2\times S^3$. The cone over $T^{1,1}$
is a singular Calabi-Yau space known as the conifold \cite{cd}; it is
described by the following equation in
${\bf C}^4$:
\begin{equation} \label{conifoldA}
\sum_{n=1}^4 w_n^2 = 0\ .
\end{equation}
When $N$ regular D3-branes are placed at the
singularity, the resulting ${\cal N}=1$ superconformal field theory
has gauge group $\suN \times \suN$.  It contains chiral superfields
$A_1, A_2$ transforming as $({\bf N},{\bf\overline N})$ and
superfields $B_1,B_2$ transforming as $({\bf\overline N},{\bf N})$,
with superpotential ${\cal W} =\lambda \epsilon^{ij} \epsilon^{kl}{\rm
Tr} A_iB_kA_jB_l$.  The two gauge couplings do not flow and 
can be varied continuously without ruining conformal 
invariance \cite{KW,MP}.
The type IIB background dual to this gauge theory is the near-horizon 
region of the solution describing $N$ D3-branes at the apex of the
conifold, namely, $AdS_5\times T^{1,1}$ with $N$ units of
5-form flux.

Adding $M$ fractional D3-branes, i.e. $M$ wrapped D5-branes, changes
the ${\cal N}=1$ gauge theory to $\suNM \times \suN$ coupled to
the bifundamental chiral superfields $A_i,\ B_j$.
This theory is no longer conformal: it can be easily seen 
that the NSVZ beta function for $g_1^{-2} - g_2^{-2}$
does not vanish \cite{GK,KN}.
Supergravity solutions corresponding to $N$ regular and $M$ fractional
branes were considered in \cite{KN,KT}.
The $M$ wrapped D5-branes introduce $M$ units of 3-form RR flux through
the 3-cycle of $T^{1,1}$:
\be \label{harmonic}
F_3 = Q \omega_3
\ ,
\ee
where $Q \sim g_s M$ and 
$\omega_3$ is the harmonic 3-form on $T^{1,1}$.
The 10-d metric is \cite{KT}
\[
ds_{10}^2 = H(r)^{-1/2} \eta_{\alpha \beta} dx^\alpha dx^\beta +
H(r)^{1/2} ( dr^2 + r^2 ds^2_{T^{1,1}} )
\]
where $\alpha,\beta =0,1,2,3$ and 
\be \label{sing}
H(r) = {Q^2 \ln (r/r_*)\over 4 r^4}
\ .
\ee
A surprising feature of the solution found in \cite{KT} is
that the 5-form flux, which corresponds to the regular D3-branes,
is not constant; in fact,
it varies logarithmically with $r$. This presence of an
extra logarithm in the metric and $\tilde F_5$ constitutes a new type
of UV behavior. Its gauge theory interpretation in terms of a cascade
of Seiberg dualities was given in \cite{KS}.

The solution of \cite{KT} is smooth at large $r$ but possesses a naked
singularity in the IR, at $r=r_*$.
In \cite{KS} this problem was removed by replacing the singular
conifold (\ref{conifoldA}) 
by the deformed conifold 
\begin{equation} \label{defconifold}
\sum_{n=1}^4 w_n^2 = \epsilon^2\ .
\end{equation}
In the deformed conifold the 2-sphere shrinks at the apex $\tau=0$, but
the 3-sphere does not. Hence the conserved 3-form flux does not
lead to a singularity of the metric; the warp factor $H(\tau)$
approaches a constant at $\tau =0$. This implies that the dual gauge
theory is confining \cite{KS}.
The deformation (\ref{defconifold})
breaks the $\uone_R$ symmetry $w_n\rightarrow
w_n e^{i\alpha}$ down to the $Z_2$ generated by $w_n\rightarrow - w_n$,
geometrically realizing the chiral symmetry breaking
in the dual gauge theory.\footnote{To be more precise, the $\uone_R$
is actually first broken to $Z_{2M}$ by instanton effects. For large $M$,
however, the  $Z_{2M}$ is well approximated by the $\uone$.}
Another interesting solution with the same asymptotics as those of
\cite{KT} was found by Pando Zayas and Tseytlin \cite{PT}.
This solution is based on the resolved conifold. It is singular in the
IR but presumably the singularity may be resolved through the enhan\c con
mechanism of \cite{JPP}. The solution of \cite{KS} describes the baryonic
branch of the $\suNM \times \suN$
gauge theory where certain baryon operators acquire expectation
values in such a way that the 
global $\sutwo \times \sutwo$ 
symmetry is unbroken while the baryon number 
$\uone_B$ is broken \cite{KS,Ahar}.
On the other hand, as conjectured in \cite{Ahar}, appropriately
resolved solution
of \cite{PT} presumably describes a mesonic branch 
of the same gauge theory where meson operators
$\tr (A_i B_j)^k$ acquire expectation values.

In this paper we consider more general solutions which
describe D$p$-branes on $R^{p,1}\times Y_{9-p}$
where $Y_{9-p}$ is a Ricci flat $(9-p)$-dimensional cone.
We consider only space-filling branes: those that fill all the
dimensions of $R^{p,1}$.\footnote{Gravity duals of fractional branes 
which are not space-filling were considered for orbifold models in
\cite{KN,Polch,Bert}.} 
The extremal background corresponding to a stack of regular branes
at the apex of the cone is well-known \cite{HorStrom}:
\be  \label{regular}
ds^2 = H(r)^{-1/2} \eta_{\alpha \beta} dx^\alpha dx^\beta +
H(r)^{1/2} ( dr^2 + r^2 h_{ij} dx^i dx^j ) \, .
\ee
We let $\alpha = 0,1, \ldots, p$
and $h_{ij} dx^i dx^j$ denotes the metric on 
$X_{8-p}$, the base of the cone.  
The dilaton and the $(p+2)$-form RR field strength
are given by 
\be
e^{4\Phi} = H(r)^{3-p}\ ,\qquad
F_{0\ldots p r} = \partial_r H^{-1}
\ .
\ee
It is interesting to ask how this background is modified by
adding a large number $M$ of D$(p+2)$-branes wrapped over a 2-cycle of
$X_{8-p}$. For this question to make sense,
we have to require that $H_2 (X_{8-p})$ 
is at least as big as ${\bf Z}$.
This is a stringent requirement: for some $p$ we will be unable to find
any supersymmetry preserving smooth spaces of this type. 

The solutions we construct in this paper are 
analogues of the Klebanov-Tseytlin (KT)
solution for $p=3$ \cite{KT}: the transverse space is taken to be 
conical. Such solutions are typically smooth in the UV but posses
a naked singularity in the IR. We will assume that the
naked singularity may be resolved by an appropriate smoothing of the cone,
but we leave construction of such solutions for the future.

Our work is closely related to the very nice papers
\cite{Cvetic1,Cvetic2} where the effect of adding RR field strength 
$F_{6-p}$ on various $p$-brane solutions was investigated.
In order for this field to be related to $M$ fractional D$p$-branes
we require that it carries $M$ units of conserved flux.
This is the generalization of the requirement (\ref{harmonic}) for
$p=3$. In \cite{Cvetic1,Cvetic2}, however, the asymptotic form
of $F_{6-p}$ 
(or of the magnetic components of $F_4$ in the case of M2-branes)
is typically such that it 
does not produce any new flux at infinity. We believe that
such solutions should not be interpreted as gravity
duals of fractional branes.
The presence of fractional branes affects the rank of the gauge group
itself. In the dual gravity description this shift of the rank
manifests itself as a new flux at infinity.
Instead, some solutions in \cite{Cvetic1,Cvetic2} have a more conventional
interpretation as deformations of the field theory present
on $N$ regular branes by certain relevant operators. 
At large $r$, through the usual AdS/CFT correspondence \cite{US,EW}, 
the power law fall-off of the field strength, and hence also of the gauge
potential, determines the dimension of the operator added to the 
dual gauge theory action.
Some specific examples of this interpretation  
will be presented in section 5. 

\section{Fractional Branes in Type IIA and IIB SUGRA}

To construct our type IIA and type IIB fractional D$p$-brane solutions, we
start with a warped product (\ref{regular})
of $R^{p,1}$ flat space-time directions and a
Ricci flat, $(9-p)$-dimensional cone $Y_{9-p}$.
Since the branes are space-filling, the warp factor depends only on the
radial coordinate of the cone.  Because the cone is Ricci flat, the base
of the cone is an $(8-p)$-dimensional Einstein manifold 
$X_{8-p}$ with metric
$h_{ij}$ normalized such that $R_{ij} = (7-p) h_{ij}$.  We assume that
this Einstein manifold has a harmonic 2-form $\omega_2$, so that  
wrapping a
D$(p+2)$-brane around the 2-cycle corresponding to $\omega_2$, and
letting the remaining $p+1$ dimensions fill
$R^{p,1}$, creates a fractional D$p$-brane.
The $\omega_2$ may be
normalized such that $\omega_2 \wedge {*}_{8-p} \; \omega_2$
is the volume form on $X_{8-p}$.

An important issue is whether such smooth manifolds $X_{8-p}$
exist and,
if so, whether they preserve any supersymmetry.
In 5 and 7 dimensions 
(corresponding to fractional D3- and D1-branes), we know of examples that
preserve some supersymmetry.  For D3-branes, the manifold $T^{1,1}$ is
such an Einstein space \cite{KT}.  For D1-branes supersymmetric
examples include $N^{0,1,0}$, $Q^{1,1,1}$ and $M^{1,1,1}$ which we
discuss in Section 3.  In the D0-, D2-, and D4-brane cases, we know
of some
Einstein manifolds with the requisite number of dimensions
and $b_2>0$: for
example, the $S^2 \times S^{6-p}$.\footnote{
After the original version of this paper was completed, 
two supersymmetric fractional D2-brane solutions were found in
\cite{Cvetic3}.  In one example, the base of the cone is
topologically $S^2 \times S^4$, while geometrically
it is an $S^2$ bundle over $S^4$.
}
However, such a
manifold does not preserve any supersymmetry and 
the resulting solution may be unstable. It is certainly unstable in
the absence of fractional branes. It would be very interesting
if there are
Einstein spaces which do not preserve supersymmetry, but which remain
stable because of the extra flux from the wrapped D$(p+2)$-branes.
We leave such issues of stability for future work.

\begin{equation}
\renewcommand{\arraystretch}{1.5}
\begin{array}{|c|ll|}
\hline
p & p\mbox{-brane field strength} & (p+2)\mbox{-brane field strength} 
\\ \hline \hline
0 & F_2 = dt \wedge dH^{-1} & 
\tilde F_4 = \frac{Q}{H r^4} dt \wedge dr \wedge \omega_2 \\ \hline
1 & \tilde F_3 = d^2x \wedge dH^{-1} &
\tilde F_5 = -Q \left( \omega_5  + {*}\omega_5 \right) \\ \hline
2 & \tilde F_4 = d^3x \wedge dH^{-1} + Q\omega_4 &
\\ \hline
3 & \tilde F_5 = d^4x \wedge dH^{-1} + {*}d^4x \wedge dH^{-1} & 
\tilde F_3 = Q \omega_3\\ \hline
4 & \tilde F_4 = H' r^4 \omega_2 \wedge \omega_2' & 
F_2 = Q \omega_2'\\ \hline
5 & \tilde F_3 = -H' r^3 \omega_2 \wedge \omega_1 & 
dC = Q \omega_1\\ \hline
\end{array}
\label{Fforms}
\end{equation}

Surprisingly, introducing $M$ wrapped D$(p+2)$-branes changes the regular
D$p$-brane solution (\ref{regular}) in a controlled way. 
We account for the D$(p+2)$-brane charge by giving a nonzero
value to $F_{6-p}$, or, for $p=0$, a nonzero value to $F_4$.  The precise
values are given in the third column of (\ref{Fforms}).\footnote{
In the table, 
$\omega_{6-p} = (-1)^p {*_{8-p}} \, \omega_2$ where the $-1$ has been added
to conform with the conventions of \cite{KS}.  For $p=4$, 
$\omega_2' = {*_4} \, \omega_2$.}
In addition, the NS-NS field must be non-zero due to the effect of
the Chern-Simons (CS) terms in the SUGRA action:
\begin{equation}
B_2 = Q {r^{p-3}\over p-3} \omega_2
\ , \qquad H_3 = d B_2= \frac{Q}{r^{4-p}} dr \wedge \omega_2
\end{equation} 
where $Q \sim g_s M$.
In the usual
D$p$-brane solution, the warp factor $H(r)$ is an eigenfunction of the
Laplacian on the Ricci flat cone with zero eigenvalue.  By adding
fractional D$p$-branes, we introduce a source term to this differential
equation:
\[
(r^{8-p} H(r)')' = -\frac{Q^2}{r^{4-p}} \ .
\]

As has become customary in gauge/gravity duality,
we integrate this equation with the boundary condition that $H$ approaches
zero as $r\rightarrow \infty$. This boundary condition
removes the asymptotically flat region
so that we ``zoom in'' on the low-energy dynamics of the dual gauge
theory.
Thus, we find that
for $p \neq 3, 5$ the warp factor is 
\begin{equation}
H(r) = \left( \frac{\rho}{r} \right)^{7-p} - 
\frac{Q^2}{(3-p)(10-2p)r^{10-2p}}
\end{equation}
where as usual $\rho^{7-p} \sim g_s N$ and $N$ 
is the number of ordinary D$p$-branes.
In the fractional D3-brane case, the warp factor takes the form familiar
from \cite{KT},
\begin{equation}
\label{logwarp}
H(r) = \left( \frac{\rho}{r} \right)^4 + 
Q^2 \left(\frac{\ln(r)}{4 r^4} + \frac{1}{16r^4} \right) \; .
\end{equation}
In the fractional D5-brane case, the warp factor is
\[
H(r) = \left( \frac{\rho}{r} \right)^2 - \frac{1}{2} Q^2 \ln r \ .
\]
Clearly, $p$ cannot exceed $5$:
for $p=6$, the
base of the cone can only be two dimensional, and there is no place to
which the flux from the wrapped D8-brane can escape.
The case $p=5$ may also be unphysical as we discuss below.

In the far UV, i.e. at large $r$, the warp factor of the D2-,
D1-, and D0-brane solutions approaches that of its $M=0$
counterpart.  In the fractional D3-brane case, the logarithmic running of
the warp factor was related to a renormalization group flow of the gauge
theory dual, and in particular to a logarithmic increase in the number of
colors in the theory \cite{KN,KT,KS}.  Curiously, the warp factor for the
fractional D4-brane solution appears to be indistinguishable from the
D5-brane warp factor in the far UV.

The case $p=5$ corresponds to the wrapped
D7-branes.  It is well known that in ten flat space-time dimensions the
D7-brane solution behaves analogously to a cosmic string in four
dimensions \cite{Green}:  the back reaction from the D7-brane makes the
metric, at least close to the D7-brane, quite complicated.  However, in
our situation, the wrapped D7-brane appears no more badly behaved than a
D5-brane at close distance, while far away in the UV there is a
logarithmic divergence which leads to a naked singularity.
This UV behavior seems rather pathological. In reality the $p=5$ solution
probably does not exist because there are
no requisite smooth three dimensional Einstein spaces.  There is a
theorem due to Hamilton \cite{Ham} which states that the only three
dimensional Einstein space with positive Ricci scalar curvature is either
$S^3$ or a quotient of $S^3$ by a discrete group.  The cone
over such a space has no
harmonic one or three cycles although it may admit vanishing two cycles.

In the infrared, the fractional D0- through D3-brane solutions exhibit
naked singularities.  In the case of the D3-brane, the
naked singularity can be resolved in one of two ways:  (1) 
The flux through the 3-cycle corresponding to the
harmonic 3-form $\omega_3 = (-1) *_5 \, \omega_2$ prevents 
the cycle 
from collapsing.  The cone is ``deformed'' and the
singularity is avoided \cite{KS}. This resolution preserves 
supersymmetry but breaks the chiral symmetry. (2)
The singularity may be hidden behind an event horizon \cite{Buchel}. 
The supersymmetry is now broken and the background describes
the high-temperature phase of the gauge theory where the chiral
symmetry is restored. 

We expect similar IR phenomena to take place
in the fractional D0-, D1-, and D2-brane
cases.  At finite temperature, the naked singularities may be cloaked by
event horizons.  Alternatively, at zero temperature, the flux through the
$(6-p)$-dimensional cycle corresponding to the harmonic form $\omega_{6-p}
= (-1)^p *_{8-p} \, \omega_2$ may deform the cone. 
We will return to these issues in a future publication. 

For each $p$, we will now describe the way in which the SUGRA solution
satisfies the requisite equations of motion.  We start with the type IIA
solutions.  Readers not interested in the details can skip ahead to the
next section.
It is a straightforward
although tedious matter to check that the field strengths and metrics
given above satisfy the Bianchi identities, field strength equations, and
the trace of Einstein's equation.  
In checking the trace, a useful formula is the equation for
the Ricci scalar in Einstein frame:
\be
R = H^{-(1+p)/8} \left( -\frac{p+1}{8} \frac{(r^{8-p} H')'}{r^{8-p}H}
+ \frac{(p+1)(p-3)}{32} \left( \frac{H'}{H} \right)^2 \right).
\label{Ricci}
\ee
We have also partially checked that the
field strengths satisfy each component of Einstein's equation
independently although without detailed knowledge of $\omega_2$, it is
difficult to do a complete check.  We believe that for a suitably
symmetric $\omega_2$, Einstein's equation will be fully satisfied.

\subsection{Type IIA Fractional Branes}

\subsubsection{Fractional D0-branes}

Fractional D0-branes involve a 9-dimensional Ricci flat cone over an
8-dimensional Einstein space $X_8$. 
The dilaton is determined
by $e^\Phi = H(r)^{3/4}$, and hence the metric in Einstein frame can be
written
\[
ds^2_E = g_s^{1/2}
\left[-H^{-7/8} dt^2 + H^{1/8} (dr^2 + r^2 h_{ij} dx^i dx^j)
\right] \ .
\] 
The nonzero field strengths are
\begin{eqnarray*}
F_2 &=& dt \wedge dH^{-1} \\
\tilde F_4 &=& \frac{Q}{r^4} H^{-1} dt \wedge dr \wedge \omega_2 \\
H_3 &=& \frac{Q}{r^4} dr \wedge \omega_2
\end{eqnarray*}
where $\omega_2$ is a harmonic 2-form on $X_8$.
The factor of $H^{-1}$ in $\tilde F_4$ 
is included to satisfy the equation of motion for the $\tilde F_4$ field
strength
\begin{equation}
d(e^{\Phi/2} {*}\tilde F_4) = -g_s^{1/2} F_4 \wedge H_3 \ .
\label{fourform}
\end{equation}
For this background $F_4 \wedge H_3 = 0$, and the factor of $H^{-1}$
in $\tilde F_4$ guarantees that $e^{\Phi/2} {*}\tilde F_4$ is independent
of $r$. The other nontrivial field equation is
\begin{equation}
d(e^{3\Phi/2} {*}F_2) = g_s e^{\Phi/2} H_3 \wedge {*}\tilde F_4 \ .
\label{twoform}
\end{equation}
Each side of this equation is proportional to the volume form on the 
9-d cone, and it is straightforward to check that the prefactors
agree too provided $(r^8 H')' = -Q^2/r^4$. The
equation for the NS-NS field $H_3$ is
\begin{equation}
\frac{g_s}{2} F_4 \wedge F_4 = d(e^{-\Phi} {*}H_3 + 
g_s^{1/2} e^{\Phi/2} C_1 \wedge {*}\tilde F_4) \ .
\label{NSNSthreeformA}
\end{equation}
It is satisfied because of two facts:  (1) $F_4 \wedge F_4=0$
and (2) $\tilde F_4$ has the factor of $H^{-1}$, making the right
side of the NS-NS three form equation of motion homogenous in $H$. Next, we
check the dilaton equation of motion
\begin{equation}
d{*}d\Phi = -\frac{g_s e^{-\Phi}}{2} H_3 \wedge {*} H_3 + 
\frac{3 g_s^{1/2} e^{3\Phi/2}}{4} F_2 \wedge {*} F_2 + 
\frac{g_s^{3/2} e^{\Phi/2}}{4} \tilde F_4 \wedge {*}\tilde F_4
\label{dilaton}
\end{equation}
and find again that the equation is satisfied provided $(r^8 H')' = -Q^2/r^4$.  

Finally, we check the trace of Einstein's equation.  The Ricci scalar on
this 10-dimensional space is given by (\ref{Ricci}) where $p=0$.
The trace of Einstein's equation is
\begin{equation}
R \; \mbox{Vol} = \frac{1}{2} d\Phi \wedge {*}d\Phi + 
\frac{g_s e^{-\Phi}}{4} H_3 \wedge {*}H_3 +
\frac{3 g_s^{1/2} e^{3\Phi/2}}{8} F_2 \wedge {*}F_2 +
\frac{g_s^{3/2} e^{\Phi/2}}{8} \tilde F_4 \wedge {*}\tilde F_4 \ .
\label{trace}
\end{equation}
where Vol is the ten dimensional volume form on the space, and the trace
equation is satisfied.

\subsubsection{Fractional D2-branes}

The dilaton is given by
$e^\Phi = H^{1/4}$, and the Einstein
frame metric is
\[
ds_E^2 = g_s^{1/2}
\left[
H^{-5/8} \eta_{\alpha\beta} dx^\alpha dx^\beta + H^{3/8}(dr^2 +
r^2 h_{ij} dx^i dx^j)
\right]
\]
where $\alpha,\beta=0,1,2$ and 
$h_{ij}$ is the metric on the 6-d
Einstein space $X_6$.  The nonzero field strengths are
\begin{eqnarray*}
\tilde F_4 &=& d^3x \wedge dH^{-1} + Q \, {*}_6 \, \omega_2 \\
H_3 &=& \frac{Q}{r^2} dr \wedge \omega_2.
\end{eqnarray*}
Only the RR 4-form
field strength is needed because under $\tilde F_4$ the
D2-branes are charged electrically while the wrapped D4-branes are charged
magnetically.

To verify the solution we
consider first the equation of motion for the field strength $\tilde F_4$
(\ref{fourform}).  Both sides of this equation are proportional to the
volume form on the seven dimensional cone, and the equation is satisfied
provided 
\be \label{allwell}
(r^6 H')' = -Q^2/r^2
\ .\ee  
The equation of motion for $H_3$
(\ref{NSNSthreeformA}) is satisfied essentially because $\tilde F_4$
includes a factor of $dH^{-1}$.  The dilaton equation of
motion (\ref{dilaton}) is satisfied because of (\ref{allwell}).
Last we check the trace
of Einstein's equation (\ref{trace}) using (\ref{Ricci}).

\subsubsection{Fractional D4-branes}

The dilaton is given by $e^\Phi = H^{-1/4}$,
and in Einstein frame the metric is
\[
ds_E^2 = g_s^{1/2}\left[
H^{-3/8} \eta_{\alpha\beta} dx^\alpha dx^\beta +
H^{5/8} (dr^2 + r^2 h_{ij} dx^i dx^j)
\right]
\]
where $\alpha,\beta=0,1,\ldots, 4$ and $h_{ij}$ is the metric on the 
4-d Einstein manifold $X_4$.  The field strengths are
\begin{eqnarray*}
\tilde F_4 &=& H' r^4 \omega_2 \wedge \omega_2' \\
F_2 &=& Q \omega_2'\\
H_3 &=& Q dr \wedge \omega_2
\end{eqnarray*}
where $\omega_2$ is the usual harmonic 2-form and $\omega_2' = {*}_4 \,
\omega_2$ is its dual.  To understand the structure of
$\tilde F_4$, consider the more familiar dual ${*}\tilde F_4 = -H^{1/8}
d^5x \wedge dH^{-1} $.  There remains seemingly an extra factor of
$H^{1/8}$.  To satisfy the equation of motion for the NS-NS 3-form
(\ref{NSNSthreeformA}), the $H^{1/8}$ is necessary because it cancels the
$e^{\Phi/2}$ that multiplies ${*}\tilde F_4$.  The field strength
equations (\ref{twoform}) and (\ref{fourform}) are trivial for this
solution.  However, the Bianchi identity for $\tilde F_4$, which was
trivially satisfied in the previous two examples, is nontrivial here:
\begin{equation}
d\tilde F_4 = -F_2 \wedge H_3 \ .
\label{Bianchifourform}
\end{equation}
The Bianchi identity leads to the differential equation for the warp
factor $(r^4 H')' = -Q^2$. The dilaton equation of motion (\ref{dilaton})
is also satisfied provided the warp factor obeys this
equation.  Finally, the trace of Einstein's equation, which we can check
provided we know the Ricci scalar curvature (\ref{Ricci}),
is also satisfied.

\subsection{Fractional Type IIB Branes}

\subsubsection{Fractional D1-branes}

In this case the transverse space is 8-dimensional, so it can be
a Calabi-Yau 4-fold. Such cases will be dual to $(1+1)$-dimensional
gauge theories with ${\cal N}=2$ supersymmetry, and we will discuss
some specific examples in section 4. In general, the
dilaton is given by $e^\Phi=H^{1/2}$, and the metric in Einstein frame is
\[
ds_E^2 = g_s^{1/2} \left[
H^{-3/4} \eta_{\alpha \beta} dx^\alpha dx^\beta + 
H^{1/4}(dr^2 + r^2 h_{ij} dx^i dx^j)
\right]
\]
where $\alpha,\beta=0,1$ and 
$h_{ij}$ is the metric on the 7-d
Einstein space $X_7$.  The nonzero field strengths are
\begin{eqnarray*}
\tilde F_3 &=& d^2x \wedge dH^{-1} \\
\tilde F_5 &=& - Q \left( \omega_5 + {*}\omega_5 \right)\\
&=& -Q \omega_5 -\frac{Q}{r^3} H^{-1} \, d^2x \wedge dr \wedge \omega_2 \\
H_3 &=& \frac{Q}{r^3} \, dr \wedge \omega_2 
\end{eqnarray*}
where $\omega_5 = (-1) {*}_7 \, \omega_2$.
By construction, the five form field strength obeys the self duality
constraint $\tilde F_5 = {*}\tilde F_5$.  Next we check the equation of
motion of the RR 3-form field strength
\begin{equation}
d(e^\Phi {*} \tilde F_3) = g_s F_5 \wedge H_3 \ .
\label{threeform}
\end{equation}
Both sides are proportional to the volume form on the eight dimensional
cone, and the equation is satisfied provided $(r^7 H')'=-Q^2/r^3$. 
The NS-NS 3-form field strength is
\begin{equation}
d{*}(e^{-\Phi} H_3 - C e^\Phi \tilde F_3) = -g_s F_5 \wedge F_3 \ .
\label{NSNSthreeformB}
\end{equation}
Note first that the axion $C=0$ for this solution.  Each side of the
equation is proportional to the eight form $d^2x \wedge dr \wedge
\omega_5$, and the coefficients match as well.

The two remaining equations to check are the dilaton equation of motion
and the trace of Einstein's equation
\begin{eqnarray}
d{*}d\Phi &=& e^{2\Phi} dC \wedge {*} dC - 
\frac{g_s e^{-\Phi}}{2}H_3 \wedge {*}H_3
+ \frac{g_s e^\Phi}{2} \tilde F_3 \wedge {*} \tilde F_3
\label{dilatonB}\\
R \, \mbox{Vol} &=& \frac{1}{2} d\Phi \wedge {*} d\Phi + 
\frac{e^{2\Phi}}{2}dC \wedge {*} dC +
\frac{g_s e^{-\Phi}}{4} H_3 \wedge {*} H_3 + 
\frac{g_s e^\Phi}{4}\tilde F_3 \wedge {*} \tilde F_3 \ .
\label{traceB}
\end{eqnarray}
The Ricci scalar is (\ref{Ricci}),
which ensures that (\ref{traceB}) is satisfied.

\subsubsection{Fractional D3-branes}

The dilaton is a constant which we will set
to zero.  The metric is
\[
ds^2 = g_s^{1/2} \left[
H^{-1/2} \eta_{\alpha \beta} + H^{1/2} (dr^2 + r^2 h_{ij} dx^i dx^j)
\right]
\]
where $\alpha,\beta = 0,1,2,3$ and $h_{ij}$ is the metric on the 5-d
Einstein space $X_5$. The nonzero field strengths are
\begin{eqnarray*}
\tilde F_5 &=& d^4x \wedge dH^{-1} + {*} d^4x \wedge dH^{-1} \\
&=& d^4x \wedge dH^{-1} - H' r^5 \omega_2 \wedge \omega_3 \\
\tilde F_3 &=& Q \omega_3 \\
H_3 &=& \frac{Q}{r} dr \wedge \omega_2 \; .
\end{eqnarray*}
This type of solution was worked out in detail in \cite{KT} for the 
supersymmetric example where the 6-d cone is the conifold.
The warp factor is given in (\ref{logwarp}). Even though this solution
is singular in the IR, it may be thought of as the asymptotic UV
part of the non-singular deformed solution presented in \cite{KS}.
In particular, it incorporates the cascade of Seiberg dualities
which takes place
in the dual ${\cal N}=1$ supersymmetric $\suNM \times \suN$
gauge theory.

The crucial feature of the conifold which enables us to write down 
the fractional brane 
solutions is that its base $T^{1,1}$ has $b_2=b_3=1$.
Other supersymmetry preserving cones with non-trivial 2-cycles are
certain generalized conifolds 
\cite{GNS, GVW, OhTatartwo}. For example, there is an $A_k$ series described by
\[
w_1^k + w_2^2 + w_3^2 + w_4^2=0
\ .
\]
Finding an explicit form of the metric on the base and of the forms
$\omega_2$, $\omega_3$ is an interesting challenge.

One could also consider a variety of 6-d cones which do not preserve
any supersymmetry. The simplest example is the space $T^{1,0}$
which is a product of $S^2$ and $S^3$ geometrically:
\[
h_{ij} dx^i dx^j = {1\over 4} ds_2^2 + {1\over 2} ds_3^2
\ ,
\]
where $ds_n^2$ is the metric on a unit $n$-sphere.
In this case $\omega_2\sim vol (S^2)$ and $\omega_3\sim vol(S^3)$ 
and the warp factor still has the form (\ref{logwarp}) \cite{KTun}.
The main concern about this explicit solution is whether 
it is stable. In the absence of the $F_3$ flux we would find
a supersymmetry breaking $AdS_5\times T^{1,0}$ background which is unstable. 
For example, there is a scalar mode with $m^2=-8$ in $AdS_5$
which\footnote{This result was found in collaboration with A. Tseytlin.}
is below the Breitenlohner-Freedman stability bound $m^2=-4$.
This particular mode corresponds to changing the relative volumes of
the $S^2$ and $S^3$ while keeping the overall volume of $T^{1,0}$
unchanged. It is clear, however, that adding the $F_3$ flux has a stabilizing
effect on this particular mode since it adds a term to the potential
which is sensitive to the volume of $S^3$.
It would be remarkable if the fractional D3-brane solution based on 
$T^{1,0}$ turns out to be stable. We postpone investigation of the
stability issue to a future publication.

\subsubsection{Fractional D5-branes}

The fractional D5-brane solution depends on a 3-d Einstein
manifold which might not exist.  We will thus be
brief in our description of this formal solution to the type IIB
SUGRA equations of motion.  The dilaton is given by $e^{\Phi} =
H(r)^{-1/2}$, and the Einstein frame metric is
\[
ds_E^2 = g_s^{1/2} \left[
H^{-1/4} \eta_{\alpha \beta} dx^\alpha dx^\beta + 
H^{3/4} (dr^2 + r^2 h_{ij} dx^i dx^j)
\right] \ .
\]
The nonzero field strengths are
\begin{eqnarray*}
dC &=& Q \omega_1 \\
\tilde F_3 &=& H^{1/2} \, {*} d^6x \wedge dH^{-1} \\
&=& -H' r^3 \omega_2 \wedge \omega_1 \\
H_3 &=& Qr \, dr \wedge \omega_2
\end{eqnarray*}
where $\omega_1 = (-1){*}_3 \; \omega_2$.
The factor of $H^{1/2}$ in $\tilde F_3$ is necessary in
order to satisfy the NS-NS 3-form field strength equation of motion
(\ref{NSNSthreeformB}):  the $H^{1/2}$ cancels with the $e^\Phi$.

The RR field strength equations of motion are trivial for this solution.  
On the other hand, the Bianchi identity $d\tilde F_3 = - dC \wedge H_3$
results in the relation $(r^3 H')' = -Q^2 r$ which ensures that
the dilaton equation of motion (\ref{dilatonB}) and the
trace of Einstein's equation (\ref{traceB}) are also satisfied.  

\section{An eleven dimensional SUGRA solution}

Recall that the SUGRA solution describing regular 
M2-branes at the apex of a cone $Y_8$ is given by the metric
\[
ds^2 = H(r)^{-2/3} \eta_{\mu\nu} dx^\mu dx^\nu + H(r)^{1/3} 
(dr^2 + r^2 h_{ij}dx^i dx^j),
\]
where $\mu,\nu=0,1,2$ and $h_{ij}$ is the Einstein metric
on the base of the cone $X_7$.
The 4-form field strength is $F_4 = d^3x \wedge dH^{-1}$.  The warp factor
$H(r)$ satisfies the differential equation $\Box_8 H(r) = 0$ where
$\Box_8$ is the Laplacian on $Y_8$.

As the only other branes in eleven dimensional supergravity are 
5-dimensional, in order to get fractional M2 branes, we need a 3-cycle
corresponding to a harmonic 3-form in the 7-dimensional base of
the cone.  It is on this three cycle that we wrap our M5-branes.
Wrapping M5 branes on such a harmonic three form $\omega_3$ produces the
following modification to the M2-brane solution.  We set
\[
F_4 = d^3x \wedge dH^{-1} + 
M {*}_7 \, \omega_3 -\frac{M}{r} dr \wedge \omega_3 \ .
\]

It is straightforward to verify that $F_4$ satisfies the Bianchi identity
$dF_4=0$ and the field strength equation $d{*}F_4 = F_4 \wedge F_4 / 2$.
The Ricci scalar on this space is
\[
R = -H^{-1/3} \left( \frac{1}{3} \frac{(r^7 H')'}{r^7 H} + 
\frac{1}{6} \left( \frac{H'}{H} \right)^2 \right) \ .
\]
{}From the Ricci scalar and the trace of Einstein's equation, 
we see that the additions to $F_4$
produce a source term in the equation for $H(r)$:
\[
(r^7 H(r)')' = -\frac{M^2}{r} \ .
\]
Thus
\[
H(r) = \left( \frac{\rho}{r} \right)^6 + M^2 
\left( \frac{\ln(r)}{6 r^6} + \frac{1}{36 r^6} \right).
\]

We conclude that adding  wrapped M5-branes produces a logarithmic
renormalization group flow in what would otherwise be a 3-dimensional
SCFT. This result is surprising in that
dimensional arguments suggest that the RG flow 
in 3-d field theories should be polynomial. We should keep in mind, however,
that we have not been able to find supersymmetry-preserving 8-d
cones with a harmonic 3-form. Non-supersymmetric such cones certainly exist
-- for example, a cone over $S^3\times S^4$ -- but it is unclear whether
they give stable solutions.

\section{Fractional D-strings}

In order to produce fractional D-strings in our framework, we need a 
7-d Einstein space with nonzero Betti numbers $b_2$ and $b_5$, and
this space should preferably preserve some supersymmetry.  Fortunately,
several examples are known to exist, and some have been well studied.  
For example, the manifold $N^{0,1,0}$ has $b_2=1$ \cite{Italians}, admits
an Einstein metric, and can be described by the coset $\suthree / \uone$.  
Moreover, this manifold, as it has three Killing spinors, is expected to
lead to a 2-d gauge theory on the regular and fractional D-strings with
${\mathcal N}=3$ supersymmetry \cite{classification}.

Two more examples are the well known manifolds $Q^{1,1,1}$ and
$M^{1,1,1}$.\footnote{ From a knowledge of the 
harmonic two forms described later, we were able
to check completely Einstein's equation for these two examples.} 
Both of these spaces admit two Killing spinors; hence, we
expect the gauge theory dual living on $N$ regular and $M$ fractional
D-strings to have ${\mathcal N} = 2$ supersymmetry \cite{classification}.  
In what follows, we will exhibit the harmonic 2-forms on these spaces.  
We will also review some facts about these spaces which allow us to guess
the Lie group structure of the gauge theory duals.  We start with the
simpler $Q^{1,1,1}$ which can be described as the coset manifold
\[
\frac{\sutwo \times \sutwo \times \sutwo}{\uone \times \uone}
\]
which has $\sutwo^3 \times \uone_R$ global symmetry.
The metric on $Q^{1,1,1}$ can be written \cite{PagePope1}
\[
ds^2 = \frac{1}{16} (d\psi - \sum_{i=1}^3 \cos{\theta_i} \, d\phi_i)^2
+ \frac{1}{8} \sum_{i=1}^3 
(d\theta_i^2 + \sin^2 {\theta_i} \,d\phi_i^2) \ ,
\]
which makes it clear that it is a $\uone$ bundle over 
$S^2\times S^2\times S^2$.
For $Q^{1,1,1}$, $b_2 = 2$.  Two linearly independent harmonic two forms are
\[
\omega_2(ij) = \frac{\sqrt{2}}{16} 
( \sin{\theta_i} \,d\theta_i \wedge d\phi_i - \sin{\theta_j} \, 
d\theta_j \wedge d\phi_j )
\]
where $(ij) = (12)$ or $(13)$.  We have normalized $\omega_2$ such that
$\omega_2 \wedge {*}\omega_2$ is the volume form on $Q^{1,1,1}$.

To understand the possible gauge theory dual to this SUGRA solution, we
review some possible embeddings of the cone over $Q^{1,1,1}$. The locus of
points describing this cone can be embedded in ${\bf C}^8$ \cite{OhTatar}:
\begin{eqnarray}
w_1 w_2 - w_3 w_4 &=& 0 \nonumber\\
w_1 w_2 - w_5 w_6 &=& 0 \nonumber\\
w_1 w_2 - w_7 w_8 &=& 0 \nonumber\\
w_1 w_4 - w_5 w_8 &=& 0 \ .
\label{Q111embedding}
\end{eqnarray}
Similar to what was done with the coset space $T^{1,1} = \sutwo \times
\sutwo / \uone$ \cite{KW}, we can reparametrize this system of equations
using instead the six complex variables $A_i, B_i, C_i$, $i=1,2$, and set
each $w_i$ to some different combination $A_i B_j C_k$.  In particular
\begin{eqnarray*}
w_1 = A_1 B_1 C_1 && w_2 = A_2 B_2 C_2 \\
w_3 = A_1 B_2 C_1 && w_4 = A_2 B_1 C_2 \\
w_5 = A_1 B_1 C_2 && w_6 = A_2 B_2 C_1 \\
w_7 = A_1 B_2 C_2 && w_8 = A_2 B_1 C_1 \ .
\end{eqnarray*}
However, we get the same $w_i$ if we act on the $A$, $B$, and $C$ by
\begin{eqnarray*}
&& A_j\rightarrow \lambda A_j\;\;\;\; B_k\rightarrow \lambda^{-1} B_k\\
&& A_j\rightarrow \mu A_j\;\;\;\; C_l \rightarrow \mu^{-1} C_l \;\;\;\; 
j,k,l = 1,2
\end{eqnarray*}
where $\lambda,\mu \in {\bf C}^*$.  We fix this freedom by selecting the
absolute value of $\lambda$ and $\mu$ to guarantee the two D-term
equations:
\begin{eqnarray*}
|A_1|^2 + |A_2|^2 &=& |B_1|^2 + |B_2|^2 \\
|A_1|^2 + |A_2|^2 &=& |C_1|^2 + |C_2|^2. 
\end{eqnarray*}
Further quotienting by the two $\uone$'s corresponding to the phases of
$\lambda$ and $\mu$ gives the cone over $Q^{1,1,1}$.  If we fix $|A_1|^2 +
|A_2|^2 = 1$, we get the space $Q^{1,1,1}$ itself.
$A_1$ and $A_2$ form a doublet under the first global $\sutwo$,
$B_1$ and $B_2$ under the second, and $C_1$ and $C_2$ under the third.

In \cite{Italians} (see also \cite{Ahn}) these
algebraic considerations were used to argue that M-theory on $AdS_4\times
Q^{1,1,1}$ is dual to a $\suN \times \suN \times \suN$ 
superconformal gauge theory in $2+1$ dimensions.
The $A_i$, $B_j$, and $C_k$ could then be understood as chiral
superfields in the
$({\bf N},\overline{\bf N},{\bf 1})$, 
$({\bf 1},{\bf N},\overline{\bf N})$, and 
$(\overline{\bf N},{\bf 1},{\bf N})$ representations of
the gauge group, respectively.  Similar considerations lead us to argue that
the gauge theory on $N$ regular D-strings placed at the apex of
the cone over $Q^{1,1,1}$
has the same Lie group structure 
and matter content as above, except in $1+1$ dimensions.
Adding $M$ fractional branes changes the gauge theory to
$\suNM \times \suN \times \suN$ coupled to chiral superfields 
$A_i$, $B_j$, and $C_k$ in the
$({\bf N+M},\overline{\bf N},{\bf 1})$, 
$({\bf 1},{\bf N},\overline{\bf N})$, and 
$(\overline{\bf N+M},{\bf 1},{\bf N})$ representations, respectively.

The next example is $M^{1,1,1}$ which has the coset structure
\cite{Wittenold,classification}
\[
\frac{\suthree \times \sutwo \times \uone}{\sutwo \times \uone \times \uone} 
\ .
\]
The Einstein metric on this seven dimensional space can be written
\cite{PagePope2}
\begin{eqnarray*}
ds^2 &=& \frac{1}{64}(d\tau + 3 \sin^2 {\mu} \,\sigma_3 + 
2 \cos {\theta_2} \, d\phi_2)^2 \\
&& + \frac{3}{4}(d\mu^2 + 
\frac{1}{4} \sin^2 {\mu} \, (\sigma_1^2 + \sigma_2^2 + \cos^2\mu \, \sigma_3^2)) \\
&& + \frac{1}{8}(d\theta_2^2 + \sin^2 {\theta_2} \, d\phi_2^2)
\end{eqnarray*}
where
\[
\sigma_1 = d\theta_1 \;\;\;\; \sigma_2 = \sin {\theta_1} \, d\phi_1 
\;\;\;\; \sigma_3 = (d\psi + \cos {\theta_1} \, d\phi_1) \ .
\]
The manifold has $b_2=1$ and thus admits one harmonic two form which can
be written in this basis as
\[
\omega_2 = \frac{1}{\sqrt{6}} \left( 
\frac{1}{8} \sin{\theta_2} \, d\theta_2 \wedge d\phi_2
- \frac{3}{16} \sin^2{\mu} \, \sigma_1 \wedge \sigma_2
+ \frac{3}{8} \sin{\mu} \cos \mu \, d\mu \wedge \sigma_3 \right).
\]
Again, we have chosen to normalize the harmonic two form such that
$\omega_2 \wedge {*}\omega_2$ is the volume form on $M^{1,1,1}$.

The cone over $M^{1,1,1}$ can be described as the locus of points in ${\bf
C}^5$ satisfying the D-term equation \cite{Italians}
\[
2(|U_1|^2 + |U_2|^2 + |U_3|^2) = 3(|V_1|^2 + |V_2|^2)
\] 
quotiented by the action of a $\uone$, acting on $U_i$ with charge +2 and
on $V_i$ with charge -3.  To recover the manifold $M^{1,1,1}$, we set
$|V_1|^2 + |V_2|^2 = 1$.  To express $M^{1,1,1}$ as an embedding in ${\bf
C}^p$ for some $p$, just as $Q^{1,1,1}$ was embedded in ${\bf C}^8$ above,
we need $p=30$.  As a result, we will not describe this embedding here.

{}From this type of algebraic consideration, the authors of \cite{Italians}
are able to deduce that in an M theory context, where the gauge theory
dual is three dimensional and conformal, the gauge group is $\suN \times
\suN$.  Moreover, they assert that $U_i$ and $V_j$ can be understood as
chiral superfields corresponding respectively to 
$\mbox{Sym}^2 {\bf N} \times
\mbox{Sym}^2 \overline {\bf N}$ and 
$\mbox{Sym}^3 {\bf N} \times \mbox{Sym}^3 \overline {\bf N}$
representations of the gauge group.  Here ${\bf N}$ and 
$\overline {\bf N}$ are
respectively the fundamental and antifundamental representations of
$\suN$.  Similar considerations lead us to argue that the gauge
theory on $N$ regular D-strings at the apex of the cone over $M^{1,1,1}$
has analogous structure, except in $1+1$ dimensions. Addition of
$M$ fractional branes again changes the group to 
$\suNM \times \suN$ and modifies the matter representation
accordingly.

Both for the $Q^{1,1,1}$ and for the $M^{1,1,1}$ example,
$\int B_2 \sim M/r^2$. As for the D3-brane solution, it is tempting to
interpret this as RG flow of a relative gauge coupling in the dual gauge 
theory. Another interesting effect is the radial variation of the
RR 3-form flux. If we assume that this flux measures the effective
number of regular branes, $N$, then
\be
N_{eff} (r) = N - a_0 {g_s M^2\over r^2}
\ ,
\ee
where $a_0$ is a proportionality constant. Unlike in the D3-brane
solutions of \cite{KT,KS}, $N_{eff}$ does not diverge in the UV.
This is in accord with the expectation that $(1+1)$-dimensional theories
should have weak UV but strong IR dynamics and hence should have power
law rather than logarithmic flow. It would be very interesting to find
detailed explanations of these supergravity effects in the
dual gauge theory. It is possible that the reduction in the number of
colors is due to some 2-d analogue of Seiberg duality, but a more mundane 
scenario, which was recently discussed in some 4-dimensional
examples \cite{Ahar}, is 
that the gauge symmetry is broken by the Higgs mechanism.
In order to understand the IR effects in more detail on the SUGRA side
it will be necessary to find resolved solutions without naked singularities.

\section{Fractional Branes and Generalizations}

Our type II and 11-dimensional supergravity solutions 
can be understood as special cases of the more
general ansatz presented by Cveti\v{c}, L\"{u}, Gibbons, and Pope in
\cite{Cvetic1, Cvetic2}.  
In the type II context they consider non-compact Ricci flat manifolds 
which are asymptotically conical and possess a harmonic 3-form
$\Omega_3$. 
They then set the NS-NS 3-form $H_3 \sim g_s M
\Omega_3$.  The warp factor must satisfy the differential equation
\[
\Box_{9-p} \, H \sim - (g_s M)^2 \, |\Omega_3|^2 \ .
\]
The symbol
$\Box_{9-p}$ is the Laplacian on the $(9-p)$-dimensional Ricci flat space.
To translate to our situation, note that $H_3 \sim (g_s M/ r^{4-p}) dr \wedge
\omega_2 $ is indeed harmonic when considered as a 3-form on the
$(9-p)$-dimensional cone.  However, 
we impose the additional requirement that $\Omega_3$ originates from
the $\omega_2$ which is harmonic on the
base of the cone.  Some specific
examples worked out in \cite{Cvetic1, Cvetic2} involve $\Omega_3$
which only exist on the full $(9-p)$-dimensional space.  
We believe that such 
solutions do not correspond to fractional D-branes 
although they are very interesting in their own right. 

Our 11-dimensional SUGRA solution can also be understood as a special
case of the general ansatz worked out in \cite{Cvetic1, Cvetic2} 
(for earlier work, see \cite{Becker,GVW})
where the authors consider asymptotically conical 
8-d Ricci flat manifolds with a self-dual 
or anti-self-dual harmonic 4-form $\omega_4$.  The 
field strength is then simply
\be
F_4 = d^3x \wedge dH^{-1} + M\omega_4  \ .
\ee
We restrict  
\be
\omega_4 = dr \wedge \omega_3/r + {*}_8 (dr \wedge \omega_3/r)
\ ,
\ee
where $\omega_3$ is a harmonic 3-form
on the base of the cone corresponding to the 3-cycle wrapped by the M5-branes.
\cite{Becker,GVW,Cvetic1,Cvetic2} allow for more general
kinds of harmonic 4-forms. 
Of the specific examples worked out in \cite{Cvetic1,Cvetic2}, 
three of the spaces have
$\sufour$ holonomy; hence the corresponding $(2+1)$-dimensional
field theory has ${\cal N}=2$ supersymmetry. The bases of these cones
are the well studied Einstein manifolds
$V_{5,2}$, $M^{1,1,1}$, and $Q^{1,1,1}$ \cite{classification}.
The fourth space has $Spin(7)$ holonomy corresponding to 
${\cal N}=1$ supersymmetry; here the base of the asymptotic cone is
the squashed $S^7$.
Although the detailed nature of $\omega_4$ is complicated in each of the 
examples, a little bit of power counting is useful in understanding
what happens.
We find that in the case of $V_{5,2}$, at large $r$, $\omega_4$
scales as $1/r^{4/3}$, for $M^{1,1,1}$ and $Q^{1,1,1}$
as $1/r^2$, and for the squashed $S^7$ as $1/r^{2/3}$.
We can ask if these asymptotics correspond to our 11-dimensional
solution found above.  The answer is no, and the reason is simple:
all four Einstein manifolds have
$b_3=b_4=0$.  
Hence, the solutions found in \cite{Cvetic1,Cvetic2} do not 
correspond to regular and fractional M5-branes. Instead, as
we now show, they describe $(2+1)$-dimensional CFT's perturbed by relevant
operators.

In each case, the 11-dimensional metric of the solution
can be written as
\[
ds^2 = H(r)^{-2/3} \eta^{\alpha\beta}dx^\alpha dx^\beta + H(r)^{1/3} ds_8^2
\]
where $ds_8^2$ is the metric on an 8-dimensional Ricci flat manifold.
Just as in the fractional brane cases, $\omega_4$ introduces a 
source term into the equation for the
warp factor: $\Box_8 H(r) \sim M^2 |\omega_4|^2$.  
At small $r$, $H(r)$ approaches a constant in all four examples,
which means that the solutions are non-singular and free of horizons. For
large $r$, the metric on the 8-dimensional manifold $ds_8^2
\rightarrow dr^2 + r^2 ds_7^2$, where $ds_7^2$ is 
the metric on the Einstein space $X_7$.  Expansions of 
$H(r)$ in powers of $1/r$ have the form
\be \label{warping}
H(r) = 
\left(\frac{\rho}{r} \right)^6 \left( 1 + \frac{c_1}{r^\gamma} + \ldots \right).
\ee
If $\gamma > 0$, then in the UV we recover the background
$AdS_4 \times X_7$ which is dual to a $(2+1)$-dimensional CFT. 
(To make the $AdS_4$ space completely manifest, we
also need to make the change of variables $r^2 = 1/z$.)
The leading deformation of the $AdS_4\times
X^7$ metric is encoded in the term
\be \sim z^{\gamma/2} ds_7^2
\ .
\ee
$\gamma$ is 4 for $M^{1,1,1}$ and $Q^{1,1,1}$, $8/3$ for
$V_{5,2}$ and $4/3$ for the squashed $S^7$ \cite{Cvetic1,Cvetic2}.
Specific examples of some of these CFT's were discussed in 
\cite{Italians, Stiefel}.

A perturbation of an $AdS_4 \times X_7$ background which falls off at
large $r$ can correspond either to adding a relevant operator to
the CFT action or to giving an expectation value to an operator 
\cite{BKL,KW2}.
In all four cases above, the perturbation that falls off the
slowest at large $r$ is contained in the 4-form $\omega_4$.
Asymptotically, we may write $\omega_4 = d c_3$, where the
$c_3$ is the perturbation of the 3-form potential. 
Purely angular components of the 3-form potential are dual to
pseudoscalar operators in the CFT \cite{aoy,minw}. 
Indeed, we will be able to
confirm that
the rate at which $c_3$ falls off at large $r$ is 
consistent with the addition of 
pseudoscalar operators to the action via the usual AdS/CFT relation
\cite{US,EW}
\be 
\phi(\vec{x},z) \sim z^{3-\Delta} \phi_0(\vec{x})
\ee
where $\Delta$ is the dimension of the operator corresponding to $\phi$ in
the conformal gauge theory dual. 
  
Let us recall that an ${\mathcal N}=1$   
chiral superfield of dimension $\Delta$ decomposes into a scalar 
of dimension $\Delta$, a fermion of dimension $\Delta+1/2$, and a 
pseudoscalar of dimension $\Delta+1$.  In searching for a pseudoscalar with
the right dimension, we need to find a chiral superfield in the conformal
gauge theory dual whose dimension is appropriate and protected.
We will present explicit findings for the $V_{5,2}$ and $Q^{1,1,1}$ cases.

For $V_{5,2}$, the cone can be embedded
in ${\bf C}^5$ via the equation
\[
F = \sum_{i=1}^5 w_i^2 = 0 \ .
\]
We expect to find protected operators in the conformal gauge theory dual
that correspond to monomials in the $w_i$ quotiented by the embedding
relation $F=0$.  In the proposed CFT dual, each $w_i$
contributes 2/3 to the operator
dimension \cite{Stiefel}. These dimensions can be derived from the AdS/CFT
correspondence as follows. From symmetry requirements, the Kaehler
potential on $V_{5,2}$ must take the form $K= (\sum_{i=1}^5
|w_i|^2)^\alpha$.  The Calabi-Yau 4-form
\[
\Omega = \frac{dw_1 \wedge dw_2 \wedge dw_3 \wedge dw_4}{w_5} \ .
\]
For the metric to be Ricci flat, the CY 4-form and the
Kaehler form $\omega = \partial \bar\partial K$ must obey the relation
$\bigwedge^4 \omega \sim \Omega \wedge \bar\Omega$.  Power
counting then shows that $\alpha=3/4$.  
It is natural to equate $K=r^2$ because
we are looking for a cone, and the metric component $g_{rr}$ should be of
order zero in $r$.  
We see that the eigenvalue of these monomials under
the action of $r \partial_r$ is $4k/3$ where $k$ is the total number of
$w_i$ in the monomial. Finally, because we need to make the transformation
$r^2= 1/z$ to make the $AdS_4$ manifest, we see that these monomials will
have dimension $2k/3$ in the conformal theory and each $w_i$ has dimension
2/3.

Now, a monomial quadratic in the $w_i$ is a
chiral field with dimension
4/3.  Also in this chiral multiplet is a pseudoscalar of dimension 7/3.
Adding this operator to the action corresponds to
$c_3\sim z^{2/3}$, precisely the rate at which $c_3$
falls off at small $z$. This perturbation breaks the conformal invariance
of the CFT dual to $AdS_4\times V_{5,2}$ and, according to the
solution found in \cite{Cvetic2}, produces confinement far in the IR.

Our next example, the cone over $Q^{1,1,1}$, can be embedded in ${\bf
C}^8$ via (\ref{Q111embedding}).
We expect to find protected operators in the gauge theory dual
which correspond to monomials in the $w_i$ quotiented by the 
embedding relations.  From \cite{Italians}, we know
that each of the $w_i$ contributes 1 to the operator dimension.  
Thus we expect and find \cite{Q111} a pseudoscalar with dimension 
2 within the chiral multiplet corresponding
to the $w_i$ themselves, explaining the falloff $c_3\sim z$ 
found in this case. 

Thus, both the $Q^{1,1,1}$ and the $V_{5,2}$
cases have a conveninent explanation in terms of perturbing 
the action by relevant operators. We also note that for the squashed
$S^7$, adding relevant operators to the action is
the only available interpretation. Here $c_3
\sim z^{1/3}$ which cannot be interpreted as due to an expectation
value; $\Delta=1/3$ required for this interpretation violates the
unitarity bound in $2+1$ dimensions, $\Delta\geq 1/2$.
Instead, the fall-off in $c_3$ 
corresponds to adding an operator of dimension $8/3$
to the action.
An analogous explanation of the $M^{1,1,1}$ case requires the
existence of a pseudoscalar operator with dimension $2$. However, we have
not found such an operator in the results of \cite{M111}.
We hope to return to this issue in the future.

A point of similarity between 
the fractional brane solutions considered here and 
the solutions constructed in \cite{Cvetic1,Cvetic2} is that all of them
have varying flux corresponding to regular branes. It is natural to interpret
the varying flux as a reduction in the number 
of degrees of freedom as the theory
flows towards the IR. In some situations \cite{KS}
this reduction is due to Seiberg duality
while in others it is due to Higgsing \cite{Ahar}. In the case
of solutions without extra conserved fluxes, such as those in
\cite{Cvetic1,Cvetic2}, the Higgsing intepretation is more likely
to be applicable. 
It would be very interesting to understand this
in more detail.

\section*{Acknowledgments}
We are grateful to O. Aharony, S. Gubser,
M. Krasnitz, M. Rangamani, M. Strassler 
and A. Tseytlin for many useful discussions.  
The work of C.H. was
supported in part by the DoD.  The work of I.R.K. was supported in part by
the NSF grant PHY-9802484 and by the James S. McDonnell Foundation Grant
No. 91-48.

\begin{appendix}
\section{Supergravity Actions and Equations of Motion}

\subsection{Type IIA}

The bosonic piece of the IIA Einstein frame action is

\begin{eqnarray}
&& \frac{1}{2\kappa^2} \int d^{10}x (-G)^{1/2} R - 
\frac{1}{4\kappa^2} \int \biggl( d\Phi \wedge *d\Phi + g_se^{-\Phi}H_3 \wedge *H_3 
\nonumber \\
&&+ g_s^{1/2} e^{3\Phi/2}F_2 \wedge *F_2  + g_s^{3/2} e^{\Phi/2} \tilde F_4 \wedge * \tilde F_4 + g_s^2 B_2 \wedge F_4 \wedge F_4 \biggr),
\end{eqnarray}
where 
\begin{equation}
\tilde F_4 = F_4 - C_1 \wedge H_3 \;\;\; , \;\;\; F_4 = dC_3 \;\;\; ,
\;\;\; F_2 = dC_1.
\end{equation}

We define the Einstein metric
by $(G_{\mu\nu})_{\rm Einstein} = g_s^{1/2}e^{-\Phi/2}
(G_{\mu\nu})_{\rm string}$.  As a result $g_s$ appears in the action, explicitly
and also through $2\kappa^2 = (2\pi)^7 \alpha'^4 g_s^2$.

The field equations are \cite{Romans}

\begin{eqnarray}
d{*}d\Phi &=& -\frac{g_s e^{-\Phi}}{2} H_3 \wedge {*}H_3 + 
\frac{3 g_s^{1/2} e^{3\Phi/2}}{4} F_2 \wedge {*}F_2 + 
\frac{g_s^{3/2} e^{\Phi/2}}{4} \tilde F_4 \wedge {*} \tilde F_4
\nonumber \\
d(e^{3\Phi/2} {*}F_2) &=& g_s e^{\Phi/2} H_3 \wedge {*}\tilde F_4 \nonumber \\
d(e^{\Phi/2} {*}\tilde F_4) &=& -g_s^{1/2} F_4 \wedge H_3 \nonumber \\
\frac{g_s}{2} F_4 \wedge F_4 &=&  
d(e^{-\Phi} {*}H_3 + g_s^{1/2}e^{\Phi/2} C_1 \wedge {*}\tilde F_4)
\nonumber \\
R_{MN} &=& \frac{1}{2} \partial_M \Phi \partial_N \Phi 
+\frac{g_s e^{-\Phi}}{4} 
( {H_M}^{PQ} H_{NPQ} - \frac{1}{12} G_{MN} H^{PQR} H_{PQR} ) 
\nonumber \\
&& +\frac{g_s^{1/2} e^{3\Phi/2}}{2} 
( {F_M}^P F_{NP} - \frac{1}{16} G_{MN} F^{PQ} F_{PQ} )
\nonumber \\
&& + \frac{g_s^{3/2} e^{\Phi/2}}{12} ( \tilde {F_M}^{PQR} \tilde F_{NPQR} - 
\frac{3}{32} G_{MN} \tilde F^{PQRS} \tilde F_{PQRS} ) e^{\Phi/2}.
\label{IIAFE}
\end{eqnarray}
We use indices $M,N,\ldots$ in ten dimensions.  The Bianchi identities are
\[
d \tilde F_4 = -F_2 \wedge H_3 \;\;\; , \;\;\;\; dF_2 = 0.
\]

\subsection{Type IIB}

The bosonic piece of the IIB Einstein frame action \cite{BBO} is 
\begin{eqnarray}
&&\frac{1}{2\kappa^2} \int d^{10}x (-G)^{1/2} R - \frac{1}{4\kappa^2}
\int \biggl( d\Phi \wedge *d\Phi + e^{2\Phi} dC \wedge *dC +
\nonumber\\
&& g_s e^{-\Phi} H_{\it 3} \wedge * H_{\it 3} +
g_s e^{\Phi} \tilde F_{\it 3} \wedge * \tilde F_{\it 3}
+ \frac{g_s^2}{2} \tilde F_{\it 5} \wedge * \tilde F_{\it 5}
+ g_s^2 C_{\it 4} \wedge H_{\it 3} \wedge F_{\it 3}\biggr) \ ,
\end{eqnarray}
supplemented by the self-duality condition
\begin{equation}
* \tilde F_{\it 5} = \tilde F_{\it 5}\ .
\end{equation}
Here
\begin{eqnarray}
\tilde F_{\it 3} &=& F_{\it 3} - C H_{\it 3}\ ,\quad F_{\it 3}
= d C_{\it 2}\ ,
\nonumber\\
\tilde F_{\it 5} &=& F_{\it 5} - C_{\it 2} \wedge H_{\it 3}\ ,\quad F_{\it 5}
= d C_{\it 4}\ .
\end{eqnarray}
The field equations are \cite{Schwarz}
\begin{eqnarray}
d{*}d \Phi &=& 
e^{2\Phi} dC \wedge {*}dC - \frac{g_s e^{-\Phi}}{2} H_3 \wedge {*}H_3
+\frac{g_s e^\Phi}{2}\tilde F_3 \wedge {*} \tilde F_3,
\nonumber \\
d( e^{2\Phi} {*}d C) &=& - g_se^\Phi H_3 \wedge {*}\tilde F_3 ,
\nonumber\\
d{*}(e^{\Phi} \tilde F_{\it 3}) &=& g_sF_{\it 5} \wedge H_{\it 3} \ ,
\nonumber\\
d{*}(e^{-\Phi} H_{\it 3} - C e^{\Phi} \tilde F_{\it 3}) &=&
- g_s F_{\it 5}
\wedge F_{\it 3} \ ,
\nonumber\\
d{*}\tilde F_{\it 5} &=& - F_{\it 3} \wedge H_{\it 3}\ ,
\nonumber\\
 R_{MN} &=& \frac{1}{2} \partial_M \Phi \partial_N \Phi +
\frac{e^{2\Phi}}{2} \partial_M C \partial_N C + \frac{g_s^2}{96}
\tilde F_{MPQRS} \tilde F_N{}^{PQRS}
\nonumber\\
&& +\frac{g_s}{4}(e^{-\Phi} H_{MPQ} H_N{}^{PQ} +
e^{\Phi} \tilde F_{MPQ} \tilde F_N{}^{PQ}) \nonumber\\
&&
- \frac{g_s}{48} G_{MN} (e^{-\Phi} H_{PQR} H^{PQR} +
 e^{\Phi} \tilde F_{PQR} \tilde F^{PQR})\ . \label{IIBFE}
\end{eqnarray}
The Bianchi identities are
\begin{eqnarray}
d\tilde F_{\it 3} &=& - dC \wedge H_{\it 3}
\nonumber\\
d\tilde F_{\it 5} &=& - F_{\it 3} \wedge H_{\it 3}\ .
\end{eqnarray}

\subsection{M theory}

The eleven dimensional SUGRA action\cite{Cremmer} is 
\begin{equation}
\frac{1}{2\kappa_{11}^2} \int d^{11} x (-G)^{1/2} R
-
\frac{1}{4\kappa_{11}^2} \int \biggl( F_4 \wedge * F_4 +
\frac{1}{3} A_3 \wedge F_4 \wedge F_4 \biggr).
\end{equation}
The field equations are then
\begin{eqnarray}
d{*}F_4 &=& \frac{1}{2} F_4 \wedge F_4, \nonumber \\
R_{MN} &=& \frac{1}{12} 
\left({F_M}^{PQR} F_{NPQR} - \frac{1}{12}G_{MN} F^{PQRS} F_{PQRS} \right)
\end{eqnarray}
supplemented by the Bianchi identity $dF_4 = 0$.
\end{appendix}


\begin{thebibliography}{99}

\bibitem{jthroat}
J.~Maldacena, ``The Large N limit of superconformal field theories and
supergravity,'' {\it Adv. Theor. Math. Phys.} {\bf 2} (1998) 231, 
{{\tt hep-th/9711200}}.

\bibitem{US}
S.S. Gubser, I.R. Klebanov, and A.M. Polyakov, ``Gauge theory correlators
from noncritical string theory,''
{\it Phys. Lett.} {\bf B428} (1998) 105,
{{\tt hep-th/9802109}}.

\bibitem{EW}
E.~Witten, ``Anti-de Sitter space and holography,''
{\it Adv. Theor. Math. Phys.} {\bf 2} (1998) 253,
{{\tt hep-th/9802150}}.

\bi{magoo}
O. Aharony, S. Gubser, J. Maldacena, H. Ooguri and Y. Oz, 
``Large $N$ Field Theories, String Theory and Gravity,'' 
{\it Phys. Rept.} {\bf 323} (2000) 183, 
{\tt hep-th/9905111}

\bi{me}
I.R. Klebanov, ``TASI Lectures:
Introduction to the AdS/CFT Correspondence,''
{{\tt hep-th/0009139}}.

\bi{ks}
S.~Kachru and E.~Silverstein, ``4d conformal field theories and
strings on orbifolds,''
{\it Phys. Rev. Lett.} {\bf 80} (1998) 4855,
{{\tt hep-th/9802183}};
A.~Lawrence, N.~Nekrasov and C.~Vafa, ``On conformal field theories
in four dimensions,''
{\it Nucl. Phys.} {\bf B533} (1998) 199,
{{\tt hep-th/9803015}}.

\bi{Kehag}
A. Kehagias, ``New Type IIB Vacua and Their
F-Theory Interpretation,''
{\it Phys. Lett.}  {\bf B435} (1998) 337, 
{{\tt hep-th/9805131}}.

\bi{KW}
I.R. Klebanov and E. Witten,
``Superconformal field theory on threebranes 
at a Calabi-Yau singularity,'' {\it Nucl. Phys.} {\bf B536} (1998) 199,
{{\tt hep-th/9807080}}.

\bi{MP}
D. Morrison and R. Plesser,
``Non-Spherical Horizons, I,''
{\it Adv. Theor. Math. Phys.} {\bf 3} (1999) 1,  {\tt hep-th/9810201}.

\bi{gipol}
E.G.~Gimon and J.~Polchinski, ``Consistency Conditions for
Orientifolds and D Manifolds,'' {\it Phys. Rev.} {\bf D54} (1996) 1667,
{\tt hep-th/9601038}.
  
\bi{doug}
M.R. Douglas, ``Enhanced Gauge Symmetry
in M(atrix) theory,'' {\it JHEP} {\bf 007} (1997) 004,
{\tt hep-th/9612126}.

\bi{GK}
S.S. Gubser and I.R. Klebanov, ``Baryons and Domain Walls in an
{\cal N}=1 Superconformal Gauge Theory,'' {\it Phys. Rev.} {\bf D58} 
(1998) 125025, {\tt hep-th/9808075}.

\bibitem{KN} I.~R.~Klebanov and N.~Nekrasov, 
``Gravity Duals of Fractional Branes and Logarithmic RG Flow,''
{\it Nucl. Phys.} {\bf B574} (2000) 263,  
{\tt hep-th/9911096}.

\bi{cd}
P.~Candelas and X.~de la Ossa, ``Comments on Conifolds,''
{\em Nucl. Phys.} {\bf B342} (1990) 246.

\bibitem{KT} I.~R.~Klebanov and A.~A.~Tseytlin,
``Gravity Duals of Supersymmetric $\suN \times \suNM$ Gauge Theories,''
{\it Nucl. Phys.} {\bf B578} (2000) 123, 
{\tt hep-th/0002159}.

\bibitem{KS} I.~R.~Klebanov and M.~J.~Strassler, 
``Supergravity and a Confining Gauge Theory:  
Duality Cascades and $\chi$SB-Resolution of Naked Singularities,''
{\it JHEP} {\bf 0008} (2000) 052, 
{\tt hep-th/0007191}.

\bi{PT}
L.~Pando Zayas and A.~Tseytlin,
``3-branes on resolved conifold,''  
{\it JHEP} {\bf 0011}(2000) 028, {\tt hep-th/0010088}.

\bi{JPP}
C. Johnson, A. Peet and J. Polchinski,
``Gauge Theory and the Excision of Repulson Singularities,''
{\it Phys. Rev.} {\bf D61} (2000) 086001, {\tt hep-th/9911161}.

\bibitem{Ahar}
O. Aharony,
``A Note on the Holographic Interpretation of String Theory
Backgrounds with Varying Flux,'' {\tt hep-th/0101013}.

\bi{Polch}
J. Polchinski, ``${\cal N}=2$ gauge--gravity duals,''
{\tt hep-th/0011193}, 
based on the talk at Strings 2000, Ann Arbor, Michigan. 

\bi{Bert}
M. Bertolini, P. Di Vecchia, M. Frau, A. Lerda, R. Marotta
and I. Pesando, ``Fractional D-branes and their gauge duals,'' 
{\tt hep-th/0011077}.

\bibitem{HorStrom} G.~T.~Horowitz and A.~Strominger, 
``Black strings and p-branes,'' {\it Nucl. Phys.} {\bf B360} (1991) 197.

\bibitem{Cvetic1} M.~Cveti\v{c}, H. L\"{u}, and C.~N.~Pope, 
``Brane Resolution Through Transgression,'' 
{\tt hep-th/0011023}.

\bibitem{Cvetic2} M.~Cveti\v{c}, G.~W.~Gibbons, H.~L\"{u}, and C.~N.~Pope, 
``Ricci-flat Metrics, Harmonic Forms and Brane Resolutions,'' 
{\tt hep-th/0012011}.

\bibitem{Cvetic3}
M.~Cveti\v{c}, G.~W.~Gibbons, H.~L\"{u}, and C.~N.~Pope, 
``Supersymmetric non-singular fractional D2-branes and NS-NS 2-branes,''
{\tt hep-th/0101096}.
%%CITATION = HEP-TH 0101096;%%

\bibitem{Green} G.~W.~Gibbons, M.~B.~Green, M.~J.~Perry, 
``Instantons and Seven-Branes in Type IIB Superstring Theory,'' 
{\it Phys. Let.} {\bf B370} (1996) 37, {\tt hep-th/9511080}.

\bibitem{Ham} R.~S.~Hamilton, 
``Three manifolds with positive Ricci curvature,'' 
{\it J. Diff. Geom.} {\bf 17} (1982) 255.

\bibitem{Buchel} A.~Buchel, 
``Finite temperature resolution of the Klebanov-Tseytlin singularity,'' 
{\tt hep-th/0011146};
S.~S.~Gubser, C.~P.~Herzog, I.~R.~Klebanov and A.~A.~Tseytlin,
``Restoration of chiral symmetry: A supergravity perspective,''
{\tt hep-th/0102172};
%%CITATION = HEP-TH 0102172;%%
A.~Buchel, C.~P.~Herzog, I.~R.~Klebanov, L.~Pando Zayas and A.~A.~Tseytlin,
``Non-extremal gravity duals for fractional D3-branes on the conifold,''
{\tt hep-th/0102105}.
%%CITATION = HEP-TH 0102105;%%

\bi{GNS}
S. Gubser, N. Nekrasov and S. Shatashvili, 
``Generalized conifolds and four dimensional N = 1 superconformal  theories,''
{\it JHEP} {\bf 9905} (1999) 003, 
{\tt hep-th/9811230}.

\bi{GVW}
S. Gukov, C. Vafa and E. Witten, ``CFT's from Calabi-Yau four-folds,'' 
{\it Nucl. Phys.} {\bf B584} (2000) 69, 
{\tt hep-th/9906070}

\bibitem{OhTatartwo}
K.~Oh and R.~Tatar, 
``Renormalization group flows on D3 branes at an orbifolded conifold,''
{\it JHEP} {\bf 0005} (2000) 030, {\tt hep-th/0003183}.

\bi{KTun}
I.~R.~Klebanov and A.~A.~Tseytlin, unpublished.

\bibitem{Italians} D.~Fabbri, P.~Fr\'{e}, L.~Gualtieri, C.~Reina, 
A.~Tomasiello, A.~Zaffaroni, and A.~Zampa, 
``3D superconformal theories from Sasakian seven-manifolds: 
new nontrivial evidences for $AdS_4/CFT_3$,'' 
{\it Nucl. Phys.} {\bf B577} (2000) 547, 
{\tt hep-th/9907219}.

\bibitem{classification} L.~Castellani, L.~J.~Romans, and N.~P.~Warner, 
``A Classification of Compactifying Solutions for d=11 Supergravity,'' 
{\it Nucl. Phys.} {\bf B241} (1984) 429.

\bibitem{PagePope1} D.~Page and C.~N.~Pope, 
``Which Compactifications of D=11 Supergravity are Stable,'' 
{\it Phys. Let.} {\bf 144B} (1984) 346.

\bibitem{OhTatar} K.~Oh and R.~Tatar, 
``Three dimensional SCFT from M2 branes at conifold singularities,'' 
{\it JHEP} {\bf 9902} (1999) 025, 
{\tt hep-th/9810244}.

\bibitem{Ahn}
C.~Ahn,
``N = 2 SCFT and M theory on AdS(4) x Q(1,1,1),''
{\it Phys. Lett.} {\bf B466} (1999) 171,
{\tt hep-th/9908162}.

\bibitem{Wittenold} E. Witten, ``Search for a Realistic Kaluza-Klein
Theory,'' {\it Nucl. Phys.} {\bf B186} (1981) 412.

\bibitem{PagePope2} D.~Page and C.~N.~Pope, 
``Stability Analysis of Compactifications of D=11 Supergravity with 
$\suthree \times \sutwo \times \uone$ Symmetry,'' 
{\it Phys. Let.} {\bf 145B} (1984) 337; 
L.~Castellani, R.~D'Auria, and P.~Fr\'{e}, 
``$\suthree \times \sutwo \times \uone$ from D=11 Supergravity,'' 
{\it Nucl. Phys.} {\bf B239} (1984) 610.

\bi{Becker}
K. Becker and M. Becker,
``M Theory on Eight-Mainfolds,''
{\it Nucl. Phys.} {\bf B477} (1996) 155, {\tt hep-th/9605053}.

\bibitem{Stiefel} A.~Ceresole, G.~Dall'Agata, R.~D'Auria, and S.~Ferrara, 
``M-Theory on the Stiefel Manifolds and 3d Conformal Field Theories,'' 
{\it JHEP} {\bf 0003} (2000) 011, 
{\tt hep-th/9912107}.

\bibitem{BKL} 
V. Balasubramanian, P. Kraus and A. Lawrence,
``Bulk vs. Boundary Dynamics in Anti-de Sitter Spacetime,''
{{\tt hep-th/9805171}}.

\bibitem{KW2} I.~R.~Klebanov and E.~Witten, 
``AdS/CFT Correspondence and Symmetry Breaking,'' 
{\it Nucl. Phys.} {\bf B556} (1999) 89,
{\tt hep-th/9905104}.

\bi{aoy}
O. Aharony, Y. Oz and Z. Yin,
``M-theory on AdS(p) x S(11-p) and superconformal field theories,''
{\it Phys. Lett.} {\bf B430} (1998) 87,
{\tt hep-th/9803051}.

\bi{minw}
S. Minwalla, 
``Particles on AdS(4/7) and primary operators on M(2/5) brane  worldvolumes,''
{\it JHEP} {\bf 9810} (1998) 002,
{\tt hep-th/9803053}.

\bibitem{Q111} P.~Merlatti, 
``M-theory on $AdS_4 \times Q^{111}$: the complete $Osp(2|4) 
\times \sutwo \times \sutwo \times \sutwo$ spectrum from harmonic analysis,''
{\tt hep-th/0012159}.

\bibitem{M111} D.~Fabbri, P.~Fr\'{e}, L.~Gualtieri, and P.~Termonia, 
``M-theory on $AdS_4 \times M^{111}$: 
the complete $Osp(2|4) \times \suthree \times \sutwo$ spectrum from 
harmonic analysis,''
{\it Nucl. Phys.} {\bf B560} (1999) 617, 
{\tt hep-th/9903036}.

\bibitem{Romans} L.~J.~Romans, ``Massive N=2a Supergravity in Ten Dimensions,''
{\it Phys. Let.} {\bf 169B} (1986) 374.

\bibitem{BBO} E.~Bergshoeff, H.~J.~Boonstra, and T.~Ortin, 
``S duality and dyonic p-brane solutions in type II string theory,'' 
{\it Phys. Rev.} {\bf D53} (1996) 7206, {\tt hep-th/9508091}.

\bibitem{Schwarz} J.~H.~Schwarz, 
``Covariant Field Equations Of Chiral N=2 D = 10 Supergravity,'' 
{\it Nucl. Phys.} {\bf B226} (1983) 269.

\bibitem{Cremmer} E.~Cremmer, B.~Julia, and J.~Scherk, ``Supergravity Theory
in Eleven Dimensions,'' {\it Phys. Lett.} {\bf 76B} (1978) 409.

\end{thebibliography}
\end{document}